\documentclass[preprint2]{aastex}

\shorttitle{Are (pseudo)bulges in isolated galaxies actually primordial relics?}
\shortauthors{Fern{\'a}ndez Lorenzo et al.}

\begin{document}

\title{Are (pseudo)bulges in isolated galaxies actually primordial relics?}

\author{M. Fern{\'a}ndez Lorenzo\altaffilmark{1}, J. Sulentic\altaffilmark{1}, L. Verdes--Montenegro\altaffilmark{1}, J. Blasco--Herrera\altaffilmark{1}, M. Argudo--Fern{\'a}ndez\altaffilmark{1}, J. Garrido\altaffilmark{1}, P. Ram{\'i}rez--Moreta, J. E. Ruiz\altaffilmark{1}, S. S{\'a}nchez--Exp{\'o}sito\altaffilmark{1} and J.D. Santander--Vela\altaffilmark{1}}
\affil{Instituto de Astrof{\'i}sica de Andaluc{\'i}a, Granada, Spain}
\email{mirian@iaa.es}

\begin{abstract}

We present structural parameters and (g--i) bulge/disk colors for a large sample (189) of isolated AMIGA galaxies. The structural parameters of bulges were derived from the 2D bulge/disk/bar decomposition of SDSS i--band images using {\tt GALFIT}. Galaxies were separated between classical bulges (n$_{b}$$>$2.5) and pseudobulges (n$_{b}$$<$2.5), resulting in a dominant pseudobulge population (94\%) with only 12 classical bulges. In the $<$$\mu_e$$>$--R$_{\rm e}$ plane, pseudobulges are distributed below the elliptical relation (smaller R$_{\rm e}$ and fainter $\mu_e$), with the closest region to the Kormendy relation populated by those pseudobulges with larger values of B/T. We derived (g--i) bulge colors using aperture photometry and find that pseudobulges show median colors (g--i)$_{b}$$\sim$1.06, while their associated disks are much bluer, (g--i)$_{d}$$\sim$0.77. Moreover, 64\% (113/177) of pseudobulges follow the red sequence of early--type galaxies. Bluer pseudobulges tend to be located in galaxies with the highest likelihood of tidal perturbation. The red bulge colors and low B/T values for AMIGA isolated galaxies are consistent with an early formation epoch and not much subsequent growth. Properties of bulges in isolated galaxies contrast with a picture where pseudobulges grow continuosly via star formation. They also suggest that environment could be playing a role in rejuvenating the pseudobulges.

\end{abstract}

\keywords{galaxies: general --- galaxies: fundamental parameters --- galaxies: interactions}

\section{Introduction}

Important clues about spiral galaxy formation lie in the nature of their central bulges. There are two main stellar systems called $"$bulges$"$ in the current literature: A) \emph{classical bulges}, characterized by an old stellar population, dynamically supported by velocity dispersion, less flat than disks and featureless; B) \emph{disk--like} or \emph{pseudo--bulges} \citep{1982ApJ...256..460K}, more flattened and rotationally supported than classical bulges, contain more dust and some show recent star formation (SF) \citep{2004ARA&A..42..603K}. Pseudobulges can contain substructures such as bars, rings and spiral arms. Surface brightness profiles of pseudobulges and classical bulges have S\'ersic index \citep{1963BAAA....6...41S} n$_{b}$$\leq$2 and n$_{b}$$\geq$2 respectively with little to no overlap \citep{2008AJ....136..773F}.

Different scenarios have been proposed for the formation of classical bulges and pseudobulges. Classical bulges are thought to form, as elliptical galaxies, in a rapid and/or violent process. This picture includes both the monolithic collapse \citep{1962ApJ...136..748E} and merger scenarios \citep{1992ApJ...399..462B}. However, pseudobulges are thought to grow slowly via redistribution of disk material augmented by gas accretion and other secular processes \citep[see][for a review]{2004ARA&A..42..603K}. Giant clumps of intense SF have been observed in high redshift spiral galaxies. Gravitational instabilities have been proposed to build bulges via clump migration \citep{2008ApJ...687...59G,2014ApJ...780...57B}, but the type of bulge remains unclear.

Optical colors of galaxies reflect mainly their stellar populations. The distribution of galaxy colors in the (g$-$r) vs. (u$-$g) plane \citep{2001AJ....122.1861S} shows a clear separation into red and blue sequences, corresponding roughly to early-- (E, S0, and Sa) and late--type (Sb, Sc, and Irr) galaxies, as expected from the respective dominance of old and young stellar populations. Since pseudobulges are thought to grow from disk material, they presumably preserve some memory of their disky origin. \citet{2009ApJ...697..630F} found a SF rate in pseudobulges similar to those found in the disks of their host galaxies (not environmentally selected), interpreting this result to be consistent with pseudobulge stellar mass growth via moderate SF. In this context the stellar population in pseudobulges should be younger and bluer than in classical bulges/elliptical galaxies. The question is, of course, how much bluer? Existing bibliography on bulge colors \citep{1996AJ....111.2238P,2001ASPC..230..237G,2004ApJS..152..175M,2004AJ....127.1371K, 2009MNRAS.393.1531G}, does not target especifically isolated galaxies, which should best reflect their origins because of their minimized environmental evolutionary effects. 

This letter presents a study of bulge colors in the AMIGA \citep[Analysis of the interstellar Medium of Isolated GAlaxies,][]{2005A&A...436..443V} sample of galaxies. This sample contains predominantly ($\sim$66$\%$) late--type spirals with small bulges \citep[B/T$<$0.1;][]{2006A&A...449..937S,2008MNRAS.390..881D}, and represents a fruitful sample to explore colors of bulges for galaxies near their primordial state (minimal merger/accretion/tidal effects). Throughout this article, the concordance cosmology with ${\Omega}_{\rm \Lambda0}=0.7$, ${\Omega}_{\rm m0}=0.3$ and $\rm H_0= \rm 70 \ km \rm \ s^{-1} \rm \ Mpc^{-1}$ is assumed.

\section{Sample selection}

The AMIGA sample is based on a refinement of the Catalogue of Isolated Galaxies \citep[CIG;][]{1973AISAO...8....3K}, where two complementary isolation parameters were defined to select the most isolated galaxies in the CIG \citep{2007A&A...472..121V}: the tidal force (Q$_{kar}$) and the local number density ($\eta_k$). The isolation parameters have been recently improved for the 636 AMIGA galaxies in the Sloan Digital Sky Survey \citep[hereafter AMIGA--SDSS;][]{2013A&A...560A...9A}. 

We selected 298 spiral (T=1--8) galaxies from the complete AMIGA--SDSS sample with recession velocities V$_r$$>$1500 km/s (median $\approx$ 7000 km/s) and isolation parameters Q$_{kar}$$<$-2 and $\eta_k<$2.7, which ensures that the galaxies have been unperturbed by mayor neighbours in the last 5 Gyr. We downloaded the images of these galaxies from the SDSS--III \citep[DR8][]{2011ApJS..193...29A,2011AJ....142...31B} in gri bands. In a few cases two or more frames were combined using the {\tt IRAF} task {\tt imcombine}. Nineteen galaxies were rejected because a bad combination of the images (2) or a close saturated star (17). The final sample is composed of 279 isolated spiral galaxies.

\section{Data analysis}

 \begin{figure*}[t]
\centering
      \includegraphics[angle=0,width=8.2cm]{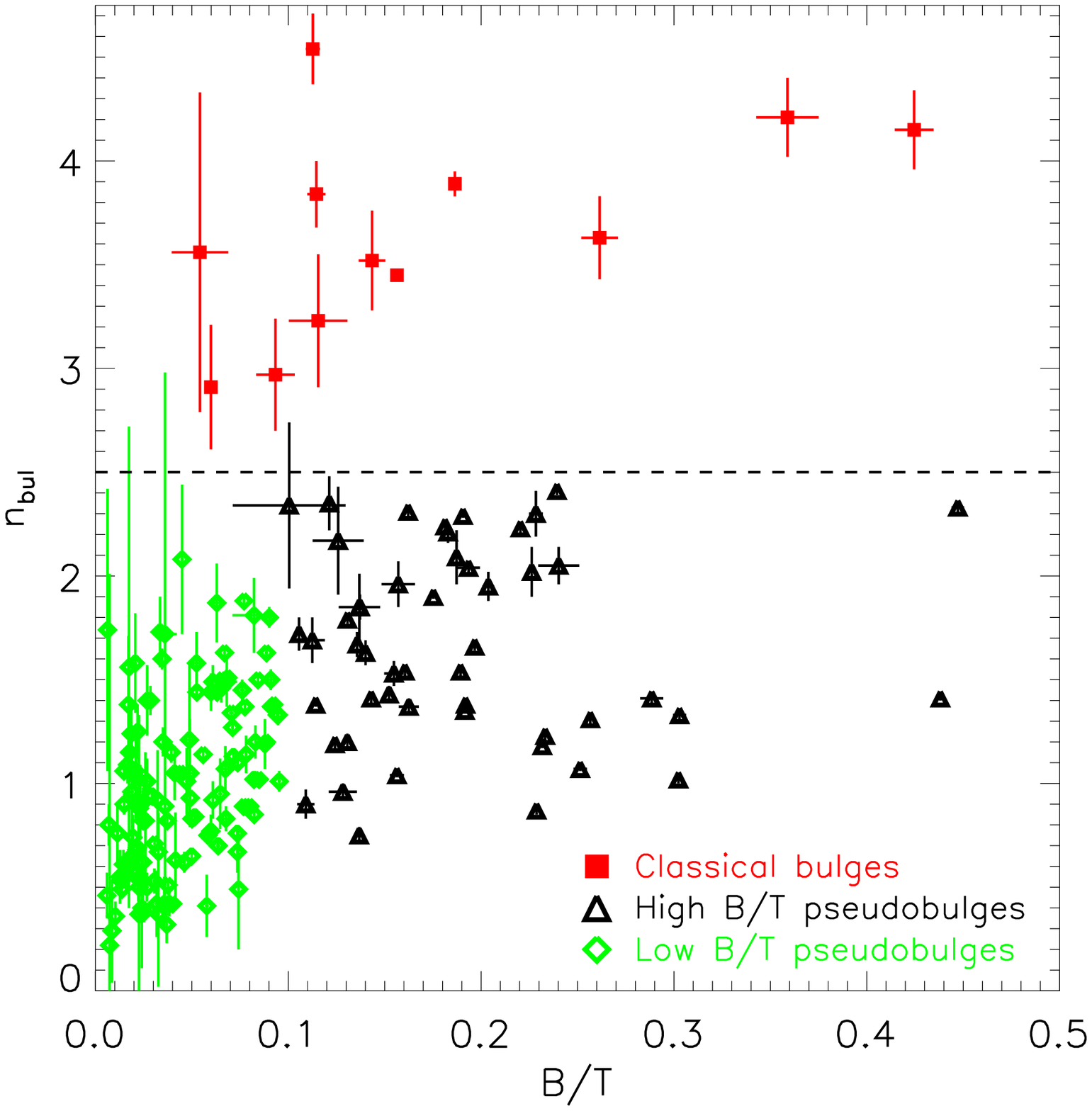}
      \includegraphics[angle=0,width=8.2cm]{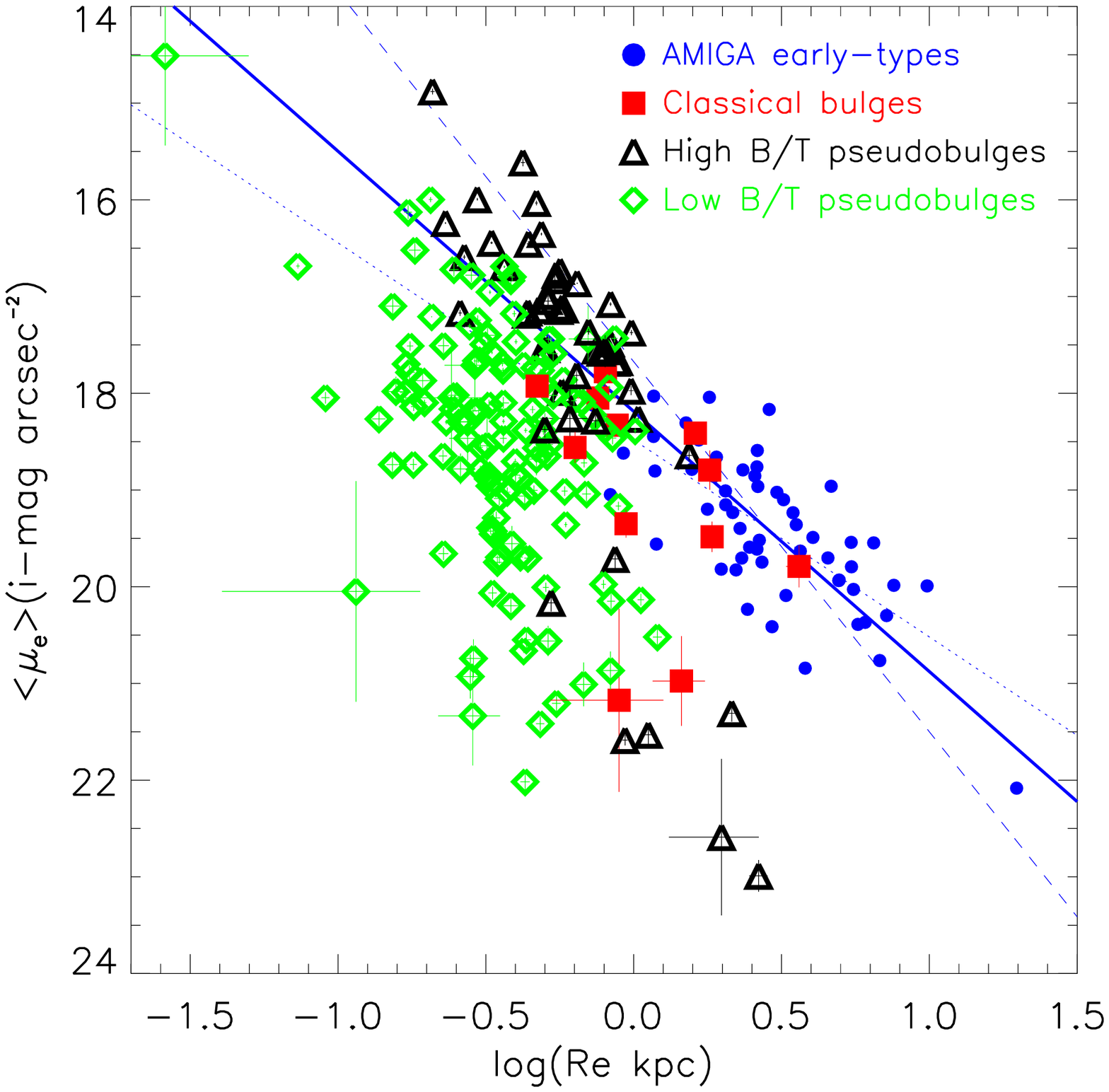}
      \caption{Structural parameters in the i--band for the bulges in the AMIGA--SDSS sample. Left panel: bulge S\'ersic index versus bulge--to--total ratio. Right panel: the Kormendy relation for classical bulges (red squares), low--B/T pseudobulges (green diamonds), and high--B/T pseudobulges (black triangles) according to the left panel. As comparison, the Kormendy relation for the AMIGA early--type galaxies is represented (blue points). A ordinary least square Y vs. X (points line), X vs. Y (dashed line) and bisector fit (solid line) to these galaxies are shown.}
 \label{fig1} 
 \end{figure*}

\subsection{Bulge structural parameters}

Total magnitudes were derived using {\tt SExtractor} \citep{1996A&AS..117..393B} in the \emph{gri} bands. The magnitudes were corrected applying the Galactic dust extinction computed by SDSS following \citet{1998ApJ...500..525S}, and the k--correction computed using the code {\tt kcorrect} \citep{2007AJ....133..734B}. 

Structural modeling of galaxy bulges was performed using the {\tt GALFIT} package \citep{2010AJ....139.2097P} applied to the i--band SDSS images. We fit a S\'ersic function to the bulges and an exponential function to the disks. We identified visually galaxies hosting a bar using the i--band and color images of SDSS and in these cases we fit an additional S\'ersic function for this component. In all cases, the model was convolved with a point--spread function (PSF) generated from the SDSS psField.

Since the fit depends strongly in the input values, we fitted each galaxy with a variety of initial values, inspecting the results and accepting a fit if the median of the residuals in the 5x5 innermost pixels was lower than 10\%. This was achieved for 189 galaxies, which form our sample hereinafter (45$\%$ barred). Four galaxies (the three Sdm and one Sd) show no bulge. Eight galaxies show signs of interaction and their disks could not be properly fitted. Four galaxies are affected by a star in the central part and 24 (most of them highly inclined) show dust lines in their centers which confuse the {\tt GALFIT} bulge fit. Finally, 50 galaxies present residuals larger than the 10\% of the flux in the innermost region, mostly caused by unreliable parameters in the fit that could not be solved with any set of initial parameters tested. In the g and r bands, the fits were done using the i--band parameters leaving free the magnitudes of the components and the disk scale lenght. We compare this metodology with the free fit in g and r--bands. The parameters obtained for the disk are very stable, with a difference in magnitude lower than 0.01 for 80\% of the cases. Differences in bulge parameters are larger although classification based in S\'ersic index (see below) would change for only 5$\%$ of the galaxies. Further discussion and details will be presented in a follow up paper.

\begin{deluxetable}{ccccccccccccccc}
\tabletypesize{\scriptsize}
\tabcolsep=0.125cm
\tablecaption{Derived data for the AMIGA--SDSS sample.}
\label{dat}
\tablehead{
{\bf CIG} & {\bf B/T} & {\bf m$_{i,d}$} & {\bf h$_R$} & {\bf (b/a)$_{d}$} & {\bf m$_{i,b}$} & {\bf a$_e$} & {\bf n$_{b}$} & {\bf (b/a)$_{b}$} & {\bf m$_{i,bar}$} & {\bf $<$$\mu_e$(i--band)$>$} & {\bf R$_e$} & {\bf M$_i$} & {\bf (g-i)$_{c,d}$} & {\bf (g-i)$_{c,b}$} \\
{} & {} &  (mag) &  (arcsec) & {} & (mag) & (arcsec) & {} & {} & (mag) & (mag/arcsec$^2$) & (kpc) & mag & {} & {} \\
{\bf (1)}  & {\bf (2)}  &  {\bf (3)} & {\bf (4)} & {\bf (5)} & {\bf (6)} & {\bf (7)} & {\bf (8)} & {\bf (9)} & {\bf (10)} & {\bf (11)} & {\bf (12)} & {\bf (13)} & {\bf (14)} & {\bf (15)} \\
}
\startdata
   1 &  0.01 & 12.75 & 10.72 &  0.52 & 17.33 &  0.99 &  0.61 &  0.56 &  {}  &  18.47 &  0.36 & -22.35 &  0.87 &  1.16 \\
   2 &  0.03 & 14.29 &  8.06 &  0.63 & 18.00 &  1.88 &  0.53 &  0.15 & 17.77 &  19.09 &  0.35 & -20.83 &  0.70 &  0.91 \\
   4 &  0.01 & 11.15 & 25.16 &  0.27 & 15.79 &  1.49 &  0.49 &  0.48 &  {}  &  17.70 &  0.17 & -21.49 &  1.08 &  1.32 \\
   7 &  0.07 & 13.98 &  6.75 &  0.73 & 16.80 &  0.51 &  1.07 &  0.79 & 17.73 &  16.84 &  0.38 & -22.45 &  0.89 &  1.17 \\
   9 &  0.04 & 14.16 &  9.14 &  0.28 & 17.66 &  1.12 &  0.42 &  0.52 &  {}  &  19.00 &  0.46 & -21.25 &  0.65 &  0.98 \\
  29 &  0.24 & 15.12 & 15.03 &  0.15 & 16.37 & 10.81 &  2.05 &  0.70 &  {}  &  22.99 &  2.65 & -19.05 &  0.63 &  0.70 \\
  33 &  0.02 & 12.64 &  9.15 &  0.60 & 16.83 &  0.89 &  0.57 &  0.85 & 15.75 &  18.30 &  0.23 & -21.21 &  0.83 &  1.08 \\
  39 &  0.07 & 13.64 &  8.99 &  0.45 & 16.53 &  2.99 &  1.46 &  0.83 &  {}  &  20.56 &  0.51 & -19.45 &  0.66 &  0.98 \\
  40 &  0.09 & 14.47 & 12.99 &  0.32 & 16.47 &  1.42 &  2.97 &  0.45 & 15.14 &  18.05 &  0.76 & -22.27 &  0.86 &  1.21 \\
  49 &  0.01 & 12.82 & 15.18 &  0.71 & 17.30 &  0.92 &  0.90 &  0.86 & 15.71 &  18.79 &  0.30 & -21.57 &  0.74 &  1.09 \\
.. & .. & .. & .. & .. & .. & .. & .. & .. & .. & .. & .. & .. & .. & .. \\
\enddata
\begin{list}{}{}
 \item[$^{\rm 1}$] Table 1 is published in its entirety in the machine-readable version and is available at http://amiga.iaa.es. A portion is shown here for guidance regarding its form and content. Columns correspond to (1): galaxy identification according to CIG catalog; (2): bulge--to--total luminosity ratio; (3): {\tt GALFIT} disk magnitude; (4) disk scalelength along the semimajor axis; (5): disk semiaxes ratio; (6): {\tt GALFIT} bulge magnitude; (7): bulge effective radius along the semimajor axis; (8): bulge S\`ersic index; (9): bulge semiaxes ratio; (10): {\tt GALFIT} bar magnitude; (11): bulge average surface brightness within the effective radius; (12): bulge effective radius; (13) total absolute magnitude in the i-band; (14) corrected disk (g--i) color; (15) corrected bulge (g--i) color. The full table contains additional data on fits and corrections.
\end{list}
\end{deluxetable}

\citet{2008MNRAS.384..420G} showed that the structural properties of bulges can be reliably retrieved when a$_e$ is larger than $\sim$80\% of the PSF half width at half maximum (HWHM). We checked the ratio between our a$_e$ (which is PSF corrected) and the seeing HWHM given by SDSS for each i--band image. Only 6 galaxies (3\% of the sample) have a$_e$/HWHM$<$0.8 and 14 galaxies have a$_e$/HWHM$<$1.

{\tt GALFIT} provides the effective radius along the semimajor axis (a$_{\rm e}$). To calculate the circular effective radius, we used R$_{\rm e}$=a$_{\rm e}$$\sqrt{(b/a)_{b}}$, where $(b/a)_{b}$ is the bulge axis ratio. We derive the average surface brightness within the effective radius using equation 1 \citep{2005PASA...22..118G}: 

\begin{equation}
<\mu_e>=m_i+2.5 \log(2 \pi R_e^2)
\end{equation}
where m$_i$ is the magnitude in the i--band given by {\tt GALFIT} for the bulge component, corrected by Galactic extinction and k--correction. $<$$\mu_e$$>$ was corrected for cosmological dimming multiplying the total flux by (1+z)$^4$, where z is de redshift of the galaxy.

 \begin{figure*}[t]
\centering
      \includegraphics[angle=0,width=16.5cm]{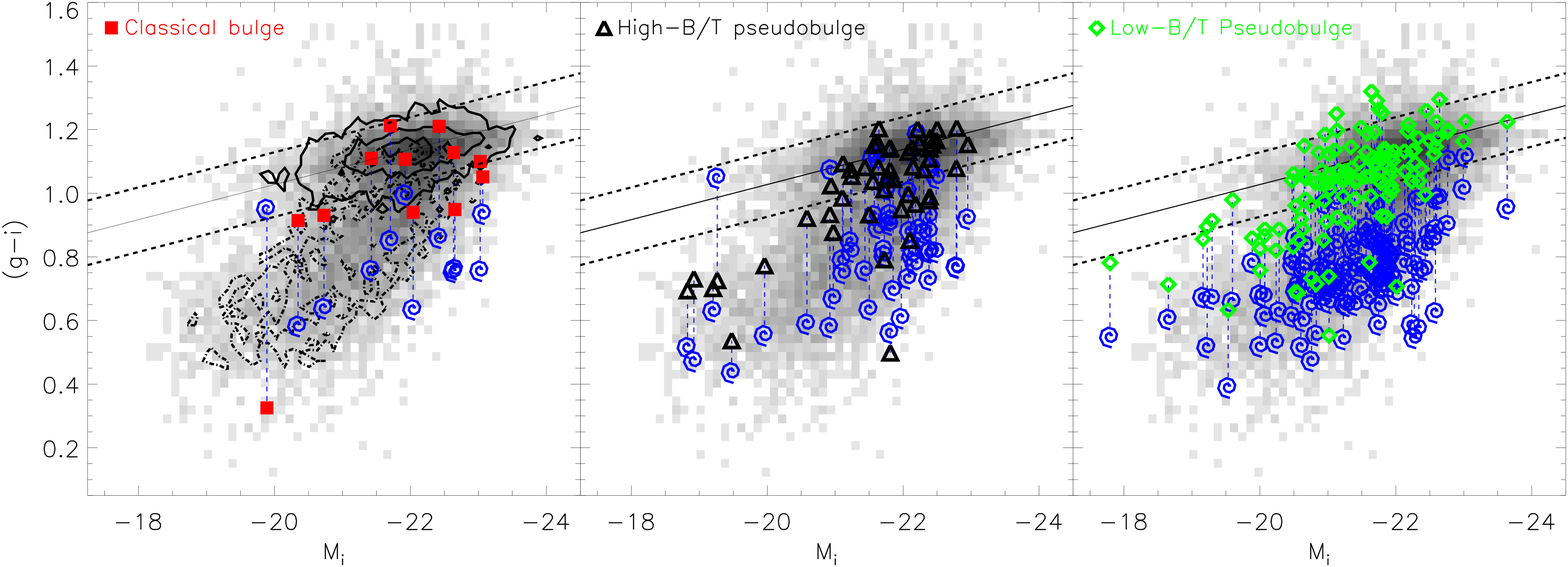}
      \caption{(g$-$i) color for bulges and the disks (blue spirals) versus the absolute magnitude in the i--band of the host galaxy (Mag$_i$). Different bulge types are represented with different symbols following Fig.~\ref{fig1}. The grey--scale represents the density diagram obtained from the \citet{2010ApJS..186..427N} sample at 0.01$<$z$<$0.05. The solid and dashed lines are the linear fit and its 2$\sigma$ for the early--type galaxies. In the first panel, we represent the contours of equal density for these galaxies (solid) and for Sbc--Sd spirals (point--dash).}
  \label{fig2}
   \end{figure*}

The bulge parameters obtained from the {\tt GALFIT} fit are represented in Fig.~\ref{fig1}. The left panel presents the bulge S\'ersic index versus the bulge--to--total (B/T) ratio.  We used n$_{b}$ to separate classical bulges (n$_{b}$$>$2.5) and pseudobulges (n$_{b}$$<$2.5). We found that 94$\%$ of bulges in the AMIGA sample are classified as pseudobulges. We used different symbols for high--B/T (B/T$>$0.1) and low--B/T (B/T$<$0.1) pseudobulges in the Figures to check if the former represent more evolved pseudobulges. In the right panel of Fig.~\ref{fig1} we show the Kormendy relation \citep{1977ApJ...218..333K} between $<$$\mu_e$$>$ and R$_{\rm e}$ for the bulges. As comparison, we represent the Kormendy relation for those early--type galaxies (T$<$0) in the AMIGA sample that were fitted with a S\'ersic function in \citet{2013MNRAS.434..325F}. To derive $<$$\mu_e$$>$, we apply the same procedure and corrections used for the bulges but m$_{i}$ is the total magnitude of the galaxy in the i--band.

Classical bulges are expected to form an unbroken sequence with the distribution of early--type galaxies in the $<$$\mu_e$$>$--R$_{\rm e}$ plane \citep{1999ApJ...523..566C}. Pseudobulges show significantly lower $<$$\mu_e$$>$ than classical bulges for a given R$_{\rm e}$ \citep{2009MNRAS.393.1531G,2010ApJ...716..942F}. The right panel of Fig.~\ref{fig1} shows results consistent with these expectations. High--B/T pseudobulges are at the top of the pseudobulges distribution, in a region consistent with the Kormendy relation, which suggests some similarity with classical bulges. \citet{2009ApJ...697..630F} found a class of non--star forming pseudobulges which are consistent with the parameter correlation of classical bulges and early--type galaxies \citep{2010ApJ...716..942F}. These inactive pseudobulges may be composite systems where both classical and pseudobulge components have similar luminosities. This could be also the case for the high--B/T pseudobulges in our sample. Moreover, some classical and pseudobulges may also be composite systems where one component is more prominent than the other \citep{2007MNRAS.379..445P,2009MNRAS.395...28M}.

\subsection{Bulge stellar populations}

We use (g--i) colors as indicative of the stellar population for bulges and disks in our sample. We used magnitudes from {\tt GALFIT} to estimate disk colors because the disk fits were very stable (see Sect. 3.1). In the case of bulges, the magnitudes derived from {\tt GALFIT} for half of the sample resulted in colors that are more than 3$\sigma$ redder than the red sequence fitted for early--type galaxies, and hence these colors are unreliable. This happens especially for later type spirals where the disk contribution is stronger. These anomalous bulge colors might be connected to a change in the disk properties inside the bulge affecting the luminosities obtained from the bulge/disk decomposition. This will be discussed in a subsequent paper. These results motivated us to derive bulge colors using aperture photometry instead. We calculated i--band aperture magnitudes for the galaxies using the {\tt IRAF} task {\tt ELLIPSE}, with fits forcing the same center for all the isophotal apertures. We then calculated the aperture magnitudes in the g and r--bands, fixing the ellipticity and position angle of each isophotal aperture equal to the i--band values. The bulge is usually defined as the excess inner light over that from the disk. Using our bulge/disk/bar decomposition, we calculate the radius where $\mu$(bulge+disk[+bar])--$\mu$(disk[+bar])=-0.1 and adopt the aperture at this radius for calculating the bulge color. This aperture is smaller than 5 pixels for 15$\%$ of the galaxies. In these cases, it was set to 5 pixels to ensure a minimum loss of bulge flux because of the seeing (width of the SDSS PSF at 1$\%$ level is $\sim$10 pixels).

 \begin{figure*}[t]
\centering
      \includegraphics[angle=0,width=16.5cm]{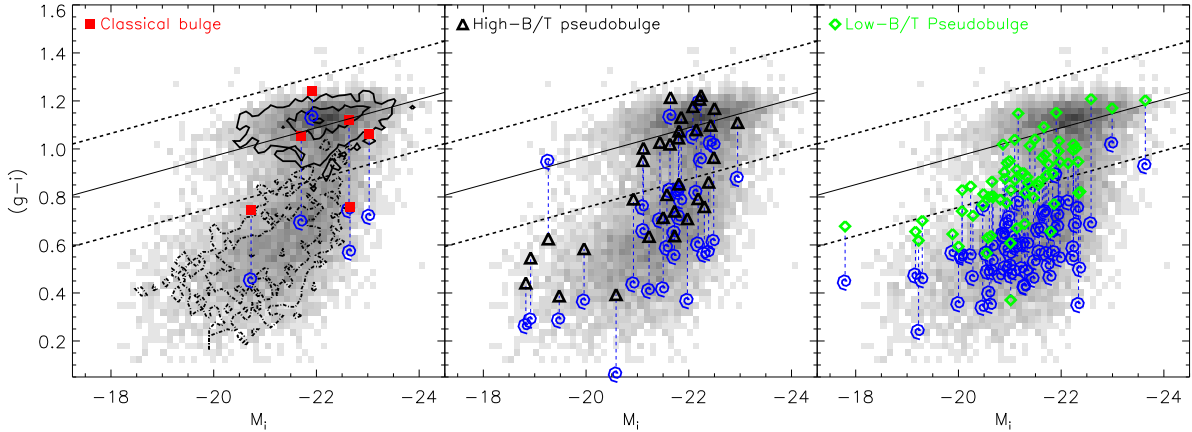}
      \caption{(g$-$i) color for bulges and disks, corrected for internal reddening, versus Mag$_i$. Symbols are the same that in Fig.~\ref{fig2}.}
  \label{fig3}
   \end{figure*}

In order to compute rest--frame colors, the magnitudes were corrected for Galactic extinction and k--correction (see Sect.3.1). We derived new k--correction for the bulge using the aperture magnitudes (the global k--corrections results in a difference in the (g--i)$_{b}$ color lower than 0.02 for 97\% of the bulges). Colors were corrected for the dependence of internal extinction with inclination following \citet[][Eq. 3]{2010MNRAS.404..792M}, but only for the Sa--Sc galaxies (87\%) since no dependence in the color with inclination was found in \citet{2012A&A...540A..47F} for later types. The same correction has been applied to bulge and disk. The structural parameters and colors calculated in this work are listed in Table 1.

In Fig.~\ref{fig2}, we present (g--i) color obtained for bulges and disks versus total i--band magnitude. We present as a comparison sample the density diagram obtained from the \citet{2010ApJS..186..427N} sample at 0.01$<$z$<$0.05 (DR8) \citep{2012A&A...540A..47F}. We find 63\% (119/189) of the bulges within 2$\sigma$ of the red sequence fitted to E/S0 galaxies of \citet{2010ApJS..186..427N} sample. Splitting by types, we find 64\% (113/177) of pseudobulges as red as early--type galaxies, with 66\% of low--B/T (84/127) and 58\% (29/50) of high--B/T pseudobulges located in the red sequence. According to the Kolmogorov--Smirnov test, the distributions of colors for low--B/T and high--B/T pseudobulges are statistically equal (probability p(K--S)=0.88), with a median (g--i) value of 1.06$\pm$0.11 and 1.07$\pm$0.13 respectively. However, the distributions of colors for their associated disks are statistically different (p(K--S)=0.003), probably because high--B/T pseudobulges are located in earlier spiral types. The mean difference between bulge and disk colors is larger for low--B/T than for high--B/T pseudobulges which likely reflects a redder disk (median (g--i)$_{d}$=0.82$\pm$0.14 versus 0.75$\pm$0.10). We found no difference between colors of classical bulges and pseudobulges (p(K--S)=0.90). This result contrasts with \citet{2009MNRAS.393.1531G}, who concludes that pseudobulges were 0.2 mag bluer than classical bulges.

The colors above are corrected only for the differential reddening due to inclination. To check if a full reddening correction could affect the results we used the extinction Av derived from the {\tt starlight} fit to the stellar continuous \citep{2005MNRAS.358..363C,2009RMxAC..35..127C} for those galaxies with SDSS spectra. This is possible for 108/189 of our galaxies and 8811/8879 of the \citet{2010ApJS..186..427N} sample. We used the \citet{2000ApJ...533..682C} law and Rv=3.1 to calculate the reddening in the (g--i) color. In Fig.~\ref{fig3}, we reproduce Fig.~\ref{fig2} with colors corrected for internal reddening. The scatter in the red sequence fitted to the E/S0 galaxies from the \citet{2010ApJS..186..427N} sample increases a factor 2 when reddening is taken into account, while the red sequence and the blue cloud are better separated. 60\% of our pseudobulges are still located in the red sequence although an important fraction of low--B/T pseudobulges seem to populate the region between the red sequence and the blue cloud, where early--type spirals and lenticulars are located.

The bulge--disk color differences found here suggest an early formation for most bulges and a different SF history for both components. Pseudobulges in the red sequence seem to follow the same color--magnitude relation than early--type galaxies, i.e., they are redder for brighter host galaxies. Also, the bluer bulges are located in the faintest spiral galaxies. These results could be indicative of a "downsizing" scenario where the most massive galaxies (and their bulges) are the first to be assembled \citep{2006A&A...453L..29C}.

\section{Discussion}

An important fraction of pseudobulges in isolated galaxies are as red as early--type galaxies. Due to the degeneracy between age and metallicity affecting the optical colors, is difficult to infer the age of this population. However, the analysis of SDSS fiber spectra for 75 AMIGA galaxies selected from \citet{2008MNRAS.390..881D} showed predominantly old stellar populations ($\sim$10 Gyr) and thus anticipated our result \citep{2012Ap&SS.337..719Z}.
 
Formation and evolution of pseudobulges are atributted to secular processes \citep{2004ARA&A..42..603K}. The young stars \citep{1996AJ....111.2238P,2001ASPC..230..237G,2009MNRAS.395...28M} and the SF found inside some pseudobulges \citep{2009ApJ...697..630F} point to this ongoing evolution. In \citet{2009ApJ...697..630F}, the specific SFR of some pseudobulges is larger than the disk specific SFR, suggesting an increase of B/T through continuous SF. 

To check whether a bulge formed through continuous SF can have a color as red as the ones observed here, we performed the following test. We did two simulations with {\tt Starburst99} \citep{1999ApJS..123....3L}, using an initial mass function of Kroupa and a metallicity of Z=0.008: 1) an instantaneous burst of 1.5x10$^9$ M$\sun$ which evolves along 8 Gyr; 2) a population which grows through a constant SF of 0.2 M$\sun$ yr$^{-1}$  (mean stellar mass of our bulges is 3x10$^9$ M$\sun$). The result after 8 Gyr is:
\begin{itemize}
\item Instantaneous burst: (g--i) = 1.13
\item Continous SF: (g--i) = 0.49
\item Combined model: (g--i) = 0.59 
\end{itemize}
Based on this, a bulge formed through continuous SF should have a very blue (g--i) color. But even if half of the mass is formed in an instantaneous burst 8 Gyr ago, the blue emission from the continuous SF dominates the color, which is also very blue in the combined model. This is still true if we assume a reddening of 0.2 in (g--i), the mean value for our bulges.

 \begin{figure}[t]
\centering
      \includegraphics[angle=0,width=7.5cm]{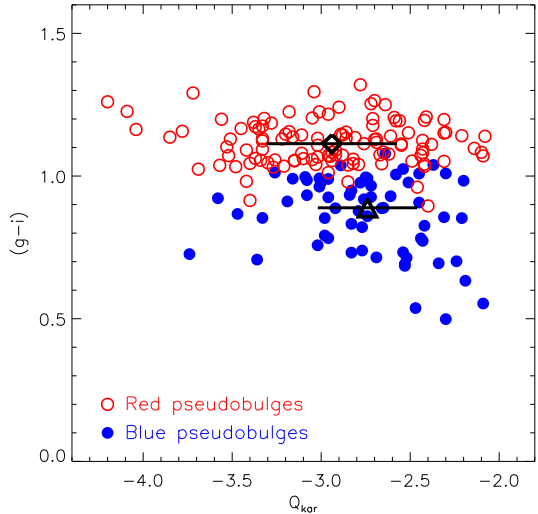}
      \caption{(g$-$i) pseudobulge color versus the tidal strength in our sample. Red open circles and blue full circles represent the pseudobulges above and below the red sequence minus 2$\sigma$ fitted in Fig.~\ref{fig2}. The median values and absolute standard deviation in Q$_{kar}$ for red (diamond) and blue (triangle) pseudobulges are shown.}
 \label{fig4} 
 \end{figure}

Gravitational instabilities \citep{2008ApJ...687...59G,2014ApJ...780...57B}, early secular evolution without much subsequent activity \citep{2004ARA&A..42..603K} or a combination of different phenomena could form pseudobulges at an early epoch. However, \citet{2009ApJ...697..630F} found strongh SF in some pseudobulges and some isolated galaxies also show blue pseudobulges colors. Could the environment play a role in rejuvenating the pseudobulges? Fig.~\ref{fig4} shows pseudobulge color versus tidal strenght Q$_{kar}$. We separated red from blue pseudobulges using as boundary the red sequence fitted in Fig.~\ref{fig2} minus 2$\sigma$. The distributions of Q$_{kar}$ for red and blue pseudobulges are statistically different (p(K--S)=0.008), with a median Q$_{kar}$ value of -2.94$\pm$0.37 and -2.74$\pm$0.28 respectively. Galaxies with red pseudobulges cover the entire range of Q$_{kar}$, while bluer pseudobulges show some preference for galaxies with larger Q$_{kar}$ (higher cross--section for environmental perturbation). This result is in agreement with \citet{2004AJ....127.1371K}, who found that galaxies with bluer bulge colors than their disks were connected with morphological peculiarities suggestive of tidal encounters.

\section{Conclusions}

We have performed a 2D bulge/disk/bar decomposition for a sample of 189 isolated galaxies. We used the S\'ersic index to separate between classical bulges (n$_{b}$$>$2.5) and pseudobulges (n$_{b}$$<$2.5). We found 12 classical bulges and 177 pseudobulges in our sample. Our pseudobulges fall below the $<$$\mu_e$$>$--R$_{\rm e}$ plane of early--type galaxies, i.e. they are less dense, with the region closest to the Kormendy relation populated by those pseudobulges with larger values of B/T.

We derived the (g--i) colors of our bulges using aperture photometry. We found that a 64\% of our pseudobulges follow the red sequence of early--type galaxies: they present colors similar to those presented by early--type galaxies as luminous as their host galaxies (redder for brighter galaxies). These red colors suggest a predominant old stellar population. The bluer bulges in our sample tend to be located in those galaxies more affected by the tidal interactions. The properties of the majority of bulges in isolated galaxies suggest that pseudobulges formed most of their mass in an early epoch, and that specific environmental events could rejuvenate the pseudobulges. In our sample of isolated galaxies, these events were minimized, which would explain our large fraction of late--types and the red colors of their (pseudo)bulges.
 
\acknowledgments

We thank the referee for his/her interest in the paper and constructive reading and suggestions. This work has been supported by Grant AYA2011-30491-C02-01 and the Junta de Andaluc{\'i}a (Spain) P08-FQM-4205 and TIC-114.

Funding for SDSS-III has been provided by the Alfred P. Sloan Foundation, the Participating Institutions, the National Science Foundation, and the U.S. Department of Energy. The SDSS-III web site is http://www.sdss3.org/.

\clearpage

\end{document}